\documentclass[graybox]{svmult}

\usepackage{mathptmx}       
\usepackage{amsmath, amssymb}
\usepackage{graphicx}
\usepackage{url}
\usepackage{tikz}
\usepackage{newtxmath}

\usetikzlibrary{arrows,positioning,shadows,shapes}
\tikzset{
   MyNodeStyle/.style={rectangle,rounded corners=1mm,draw=black,
                      top color=white, bottom color=blue!40,
                      thick,inner sep=2.8mm, xshift=-4mm, font=\scriptsize,
                      minimum size=1mm, text centered,drop shadow},
   MyArrowStyle/.style={->, >=latex', shorten >=0pt, thick},
   MyLabelStyle/.style={text centered, font=\scriptsize}}
   
\pgfdeclarelayer{background}
\pgfdeclarelayer{foreground}
\pgfsetlayers{background,main,foreground}

\def\bC{\mathbf{C}}
\def\bF{\mathbf{F}}
\def\bx{\mathbf{x}}
\def\bu{\mathbf{u}}
\def\ie{i.e.\ }
\def\eg{e.g.\ }

\begin{document}

\title{Development of an efficient and flexible pipeline for Lagrangian coherent structure computation}
\titlerunning{Development of LCS pipeline using GPGPU and VTK} 
\author{Siavash Ameli \and Yogin Desai \and Shawn C. Shadden}
\institute{Siavash Ameli, Shawn C. Shadden \\
University of California, Berkeley, email: \{sameli, shadden\}@berkeley.edu \vspace{1em} \\
Yogin Desai\\
Georgia Tech, email: yogindesai02@gmail.com}

\maketitle

\abstract{The computation of Lagrangian coherent structures (LCS) has become a standard  tool for the analysis of advective transport in unsteady flow applications. LCS identification is primarily accomplished by evaluating measures based on the finite-time Cauchy Green (CG) strain tensor over the fluid domain. Sampling the CG tensor requires the advection of large numbers of fluid tracers, which can be computationally intensive, but presents a large degree of data parallelism. Processing can be specialized to parallel computing architectures, but on the other hand, there is compelling need for robust and flexible implementations for end users.  Specifically, code that can accommodate analysis of wide-ranging fluid mechanics applications,  while using a modular structure that is easily extended or modified, and facilitates visualization is desirable.  We discuss the use of Visualization Toolkit (VTK) libraries as a foundation for object-oriented LCS computation, and how this framework can facilitate integration of LCS computation into flow visualization software such as ParaView. We also discuss the development of CUDA GPU kernels for efficient parallel spatial sampling of the flow map, including optimizing these kernels for better utilization. }

\section{Introduction} \label{sec:introduction}

The computation of Lagrangian coherent structures (LCS) originated as a means to compute stable and unstable manifold type structures in vector fields with aperiodic time dependence. This was motivated by knowledge that the interaction of such manifolds gives rise to chaotic dynamics, and hence understanding these interactions helped bring an ordered understanding to chaotic advection in fluid flow. This approach was originally applied to time periodic systems, especially using Poincar\'{e} mappings that make the dynamics autonomous. The applicability of traditional invariant manifold concepts breaks down for time aperiodic vector fields for practical and conceptual reasons. Notably, asymptotic notions associated with invariant manifold theory are neither applicable nor desirable for understanding inherently transient phenomena associated with unsteady computational or experimental fluid flow data.  

LCS computational techniques do not typically solve for the stable and unstable manifolds of explicit trajectories. A global approach was developed from observations that material points straddling stable and unstable manifolds typically separate faster in forward and backward time than pairs of points not straddling such manifolds.  That is, the manifold geometry may be inferred by considering the stretching associated with the hyperbolicity of these structures~\cite{JonesWinkler02}. This led to an alternative characterization of  organizing structures in fluid flows as invariant manifolds (material surfaces) that satisfy certain locally attracting or repelling properties, which became the basis for formalizing the concepts of LCS. LCS computations have been applied to diverse applications, as reviewed in~\cite{Shadden12}, demonstrating wide-ranging utility in analysis of unsteady fluid advection. 

Scalar measures are often used to identify LCS, such as the (largest) finite time Lyapunov exponent FTLE field~\cite{ShaddenLekienMarsden05}. This is accomplished by plotting the field and visually identifying such structures, or using an algorithmic approach to extract features such as ridges in the field~\cite{SadloPeikert07, PeikertSadlo08, SchindlerETAL12}.  The FTLE is derived from the largest eigenvalue of the Cauchy Green (CG) strain tensor
\begin{equation}\label{eq:strain-tensor}
\bC(\bx_0, t_0, t_f) = \nabla \bF_{t_0}^{t_f}(\bx_0)^{\intercal}  \cdot \nabla \bF_{t_0}^{t_f}(\bx_0) \;,
\end{equation}
where $\bF_{t_0}^{t_f}\colon \bx(t_0) \mapsto \bx(t_f)$ denotes the flow map, and $\bx(t)$ is a fluid element trajectory with $\bx_0=\bx(t_0)$. The full CG tensor encodes direction-dependent stretching information that can be leveraged, for example in defining normally hyperbolic LCS, \ie material surfaces that are locally the most normally repelling over a chosen time interval~\cite{Haller11}. Therefore, the computation of the CG tensor over the flow domain can be thought of as a common, or at least representative, target in LCS identification strategies.   

The global approach to LCS identification requires a highly resolved sampling of the CG tensor over the fluid domain to locate potential LCS. One typically starts with fluid velocity field data $\bu(\bx, t)$, obtained from computation or measurement, and the flow map is computed by seeding the fluid domain with tracers and integrating the advection equation
\begin{equation}  \label{eq:ode}
\dot{\bx}(\bx_0, t) = \bu(\bx, t) \; ,
\end{equation}
over a finite time interval $(t_0, t_f)$ for a grid of seed points $\mathbf{x}_0$. This is typically the most computationally intensive aspect of LCS identification since the seed grid may be composed of millions of material points especially in 3D flows. Furthermore, the integration of Eq.~(\ref{eq:ode}) for each seed point typically requires thousands of integration steps or more, and each integration step can require several space-time interpolations of $\bu(\bx, t)$. Furthermore, this process may be repeated for a time series of seed grids at different $t_0$ to obtain the time evolution of LCS.  We note that the high resolution of seed points is needed in part for finite differencing to compute $\nabla \bF_{t_0}^{t_f}(\bx_0)$. Alternatively, integration of the variational equations, as performed in~\cite{KastenETAL09}, may also be used to obtain the linearized flow map directly. However, sufficient resolution is still needed for LCS detection, especially when ridge extraction is performed~\cite{SchindlerETAL12}. 

It has been our experience that two major deterrents for wider adoption of LCS computations for  flow post-processing are (1) computational time and (2) ease of use. With regards to (1), we will discuss the use of general purpose computing on graphics processing units (GPGPU) for accelerating flow map computations. This work builds on previous works of~\cite{GarthETAL07, JimenezVankerschaver09, ContiRossinelliRossinelli12}. With regards to (2), we will discuss the use of the Visualization Toolkit (VTK) for creating an LCS computational pipeline that on the ``front-end'' is capable of handling various input data, and on the ``back-end'' facilitates visualization of LCS with standard flow analysis tools. In Sec.~\ref{sec:pipeline} we describe the main component of the LCS computational pipeline. In Sec.~\ref{sec:parallel} we describe the implementation of the flow map computation on the GPU and results for various applications, ranging from 2D Cartesian to 3D unstructured velocity data processing. In Sec.~\ref{sec:discussion} we will discuss some results and relationship with previous work. 

\section{LCS pipeline} \label{sec:pipeline}

Here we describe a pipeline developed to process velocity field data for the purpose of computing LCS. This pipeline was written in C++ and will be made openly available on GitHub (\url{www.github.com/FlowPhysics}). This pipeline was developed using an object-oriented approach to create filters (classes that modify data) that can be extended or modified with minimal effort. Specifically, this pipeline makes extensive use of the popular Visualization ToolKit (\url{www.vtk.org}). VTK is a broad collection of open-source libraries for 3D computer graphics, image processing and visualization. The filters that we developed adhere to VTK coding standards and conventions for modularity, portability and debugging. We also chose to develop our pipeline using VTK for two additional reasons. First, VTK has a number of classes for reading standard file formats and data grid types commonly used by the fluid mechanics community. We could therefore leverage existing functionality to better support input from a variety of applications. Second, open-source programs for visualizing scientific data such as ParaView and VisIt are built on VTK.  Developing an LCS pipeline in VTK facilitates integration with these programs, and alternatively development of custom rendering for a stand-alone software.  

VTK is highly object-oriented and supports a pipeline architecture that generalizes the connection and execution of algorithms. The benefits of this methodology are (1) a universal interface between different filters in the pipeline, (2) greater control over information flow to better manage cache. The pipeline enables streaming of data, which is important for the application here because unsteady velocity field data are typically stored by a series of files that are often altogether too large to be loaded in memory. Streaming requires the use of \emph{time-aware} filters and for this we developed time-related request keys, similar to the approach described by Biddiscombe, et al.~\cite{BiddiscombeETAL07}. This pipeline does not support spatial streaming of velocity data for applications where individual time frames are too large for memory. 

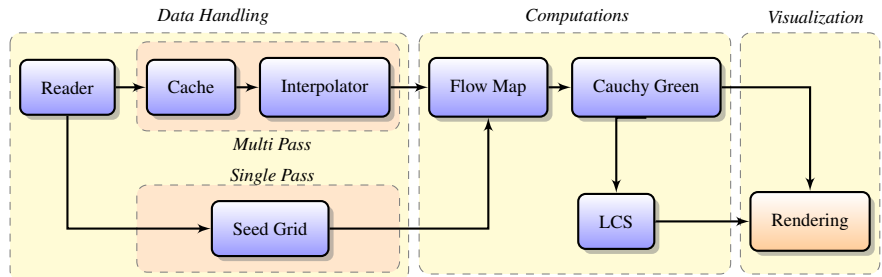
\begin{figure}
	\centering

	\begin{tikzpicture}[scale=0.4]
	
	\node [MyNodeStyle](Cache){Cache};
	\node [MyNodeStyle,right=of Cache,xshift=-0.3cm](Interpolator){Interpolator};
	\path [draw, ->,MyArrowStyle] (Cache) -- node [name=CacheToInperpolator] [above] {}
	       (Interpolator.west |- Cache) ;
	\node [MyNodeStyle,below= of CacheToInperpolator,yshift=-0.5cm,xshift=0.7cm](Seed){Seed Grid};
	\node [MyNodeStyle,left=of Cache,xshift=1cm](Reader){Reader};
	\node [MyNodeStyle,right=of Interpolator,xshift=-0.1cm](FlowMap){Flow Map};
	\node [MyNodeStyle,right=of FlowMap,xshift=-0.3cm](CauchyGreen){Cauchy Green};
	\node [MyNodeStyle,below=of CauchyGreen](LCS){LCS};
	\node [MyNodeStyle,right=of LCS, xshift=0.65cm, bottom color=orange!40]
	      (Visualization) {Rendering};
	
	\draw [MyArrowStyle] (Reader.east) -- ++(0,0) -- ++(0,0) |-  (Cache.west);
	\draw [MyArrowStyle] (Interpolator.east) -- ++(0,0) -- ++(0,0) |-  (FlowMap.west);
	\draw [MyArrowStyle] (FlowMap.east) -- ++(0,0) -- ++(0,0) |-  (CauchyGreen.west);
	\draw [MyArrowStyle] (CauchyGreen.south) -- ++(0,0) -- ++(0,0) -|  (LCS.north);
	\draw [MyArrowStyle] (Reader.south) -- ++(0,0) -- ++(0,0) |-  (Seed.west);
	\draw [MyArrowStyle] (Seed.east) -- ++(0,0) -- ++(0,0) -|  (FlowMap.south);
	\draw [MyArrowStyle] (LCS.east) -- ++(0,0) -- ++(0,0) |- (Visualization.west);
	\draw [MyArrowStyle] (CauchyGreen.east) -- ++(0,0) -- ++(0,0) -|  (Visualization.north);
	
	\begin{pgfonlayer}{background}
	
		\path (Reader.west |- Interpolator.north)+(-0.3,0.8) node (Rectangle0-Point1) {};
		\path (Seed.south -| Interpolator.east)+(+0.6,-0.8) node (Rectangle0-Point2) {};
   	 	\path[fill=yellow!20,rounded corners, draw=black!50, dashed]
   	 	      (Rectangle0-Point1) rectangle (Rectangle0-Point2) 
   	 	      node[MyLabelStyle,anchor=north,above,pos=0,xshift=2.7cm]
   	 	      (Rectangle0) {\textit{Data Handling}};
	
		\path (Cache.west |- Interpolator.north)+(-0.3,0.5) node (Rectangle1-Point1) {};
		\path (Cache.south -| Interpolator.east)+(+0.3,-0.5) node (Rectangle1-Point2) {};
    	        \draw[fill=orange!20,rounded corners, draw=black!50, dashed]
    	             (Rectangle1-Point1) rectangle (Rectangle1-Point2)
    	             node[MyLabelStyle,anchor=north,above,pos=0,xshift=1.8cm,yshift=-1.55cm]
    	             (Rectangle1) {\textit{Multi Pass}};
    
		\path (Cache.west |- Seed.north)+(-0.3,0.5) node (Rectangle2-Point1) {};
		\path (Seed.south -| Interpolator.east)+(+0.3,-0.5) node (Rectangle2-Point2) {};
    		\path[fill=orange!20,rounded corners, draw=black!50, dashed]
    	      	     (Rectangle2-Point1) rectangle (Rectangle2-Point2)
    	             node[MyLabelStyle,anchor=north,above,pos=0,xshift=1.8cm,yshift=-0.1cm]
    	             (Rectangle2) {\textit{Single Pass}};
    
		\path (FlowMap.west |- CauchyGreen.north)+(-0.3,0.8) node (Rectangle3-Point1) {};
		\path (LCS.south -| CauchyGreen.east)+(+0.3,-0.8) node (Rectangle3-Point2) {};
    	        \path[fill=yellow!20,rounded corners, draw=black!50, dashed]
    	             (Rectangle3-Point1) rectangle (Rectangle3-Point2)
    	             node[MyLabelStyle,anchor=north,above,pos=0,xshift=2.1cm]
    	             (Rectangle3) {\textit{Computations}};
    
		\path (Visualization.west |- CauchyGreen.north)+(-0.3,0.8) node (Rectangle4-Point1) {};
		\path (LCS.south -| Visualization.east)+(+0.3,-0.8) node (Rectangle4-Point2) {};
    	        \path[fill=yellow!20,rounded corners, draw=black!50, dashed]
    	        (Rectangle4-Point1) rectangle (Rectangle4-Point2)
    	        node[MyLabelStyle,anchor=north,above,pos=0,xshift=1cm]
    	        (Rectangle4) {\textit{Visualization}};
    
               \end{pgfonlayer}
    
	\end{tikzpicture}

\caption{Pipeline Architecture} \label{fig:pipeline}

\end{figure}

Our pipeline architecture includes the following custom filters, which are connected as shown in Fig.~\ref{fig:pipeline}. We note that the pipeline does not represent data flow, but rather connectivity of filters and the management of requests.

\texttt{{\bf Reader.}} A reader filter was developed to accept a variety of input data. VTK contains a number of classes to load various types of data (e.g. structured and unstructured grid data) and file formats. There are few readers in VTK that are time-aware, and those may only support one or a few file formats. Our filter implements an reader class that casts to the appropriate readers of both legacy and XML file types to support eleven file formats commonly used in VTK. This filter provides additional functionality by providing two different types of output based on requests of the pipeline. Output can be either a single data object or a data object that encapsulates multiple data objects depending on if the request comes from the \texttt{Seed} or the \texttt{Cache} filter. This filter also adds metadata on output objects such as times and indices that are needed throughout the pipeline to facilitate temporal streaming and interpolation. Lastly, this filter provides a framework for implementing necessary pre-processing of velocity data, such as tetrahedralization as motivated below.
	
\texttt{{\bf Cache.}} This filter is connected to the output of \texttt{Reader}. The \texttt{Cache} filter is essentially a wrapper that manages VTK's \texttt{vtkTemporalDataSetCache} class. It stores a window of the last requested data time frames.  The amount of data cached depends on the upstream request.  This filter is used to avoid repetitive streaming processes in the pipeline and deletes unnecessary data on memory.

\texttt{{\bf Interpolator.}} This filter is used to interpolate the velocity field data in space and time. Data in the pipeline before this filter are provided discretely and all filters after \texttt{Interpolator} treat velocity data as continuous. Tracers are updated altogether each time step. This ensures that tracers are requesting the same flow field information each update step to best handle data streaming and memory utilization. This is also consistent with our GPU implementation whereby tracers are updated altogether to keep threads short-lived. We note that both space and time interpolation is performed on the GPU. Specifically, a window of velocity data frames (nominally two velocity files for linear interpolation) are loaded on the GPU's memory, and interpolation is performed on the GPU as necessary as integration proceeds between these time frames. 

\texttt{{\bf Seed.}} 
This filter is used to initialize the seed points where the CG tensor is to be sampled. Currently these locations are defined as structured grid data. This filter defines their initial conditions, other attributes, and for unstructured velocity data computes which cell in the velocity field mesh each seed point is located in. While the seed grid are nominally the locations where the CG tensor is evaluated, highly localized auxiliary points may also be defined about each seed point for improving flow map gradient computation as discussed below. 

\texttt{{\bf Flow Map.}} This filter maps the seed points forward in time. Therefore it is connected to both \texttt{Seed} and \texttt{Interpolator}.  It is capable of implementing various types of integration routines, \eg single step (Runge Kutta) and multi step (\eg Adams Bashforth) methods. The same points passed from the input (\texttt{Seed}) are passed to the output with the final seed point locations added as a vector attribute to the point data. If a trajectory leaves the spatial domain of the velocity data, two options are possible. The CG tensor evaluation can be performed early at all locations relying on this trajectory.  Alternatively, velocity extrapolation can be performed inside the interpolation filter, however one must take care is this regard~\cite{TangChanHaller10, Shadden12}.

\texttt{{\bf Cauchy-Green.}} This filter calculates the finite time \texttt{Cauchy-Green} strain tensor over the seed points using the initial and final point locations output from \texttt{Flow Map}. Currently, this is performed using standard central difference formula for the flow map gradient matrix entries. This differencing can be applied directly to the evolution of the seed point grid, or by differencing auxiliary points attributed to each seed location as described above. The auxiliary point method is more computationally expensive, but can provide more accurate CG tensor computation at less computational expense than an equivalent increase of seed grid resolution when seed spacing is reduce by more than 1/2. However, an equivalent increase in seed grid resolution also improves resolution of the CG tensor field, which is not accomplished through the auxiliary points method. Additionally, the auxiliary point method can result in missed structures, when LCS are not properly straddled by points used in finite differencing~\cite{ShaddenLekienMarsden05, UeffingerETAL12}. This filter also computes the eigenvalues and eigenvectors of the CG tensor at each seed point, which are commonly used for LCS detection. The seed points are passed to the output with the eigenvalue/vector fields added as point data attributes.   

\texttt{{\bf LCS.}} This filter processes the output from the \texttt{Cauchy Green} filter to apply criterion for LCS detection, for example, further processing of FTLE field data, such as Gaussian smoothing to remove computational noise as well as C-Ridge computation~\cite{SchindlerETAL12, SchindlerETAL12b}. While not currently implemented, strainline and shearline computation as described in~\cite{HallerBeronVera12} could be implemented in this filter. 

\texttt{{\bf Visualization.}} This is not a separate filter in our pipeline, but we have separated this conceptually. One important benefit of using the new pipeline functionality of VTK is that we can compile each filer to a shared library that can be loaded by the ParaView plugin manager, so an end user can simply import the filters as plugins in ParaView and then use these to create a pipeline for computing and visualization of fields defined from the CG filter (e.g. FTLE), or visualize features from the LCS filter. We note that support for FTLE computation was recently added to VisIt~\cite{VisIt} as used by~\cite{OzgokmenETAL12}.

\section{Flow map computation on the GPU} \label{sec:parallel}

The computation of the CG tensor field requires the advection of a large set of tracer trajectories. Since each tracer moves independently, these calculations present a parallel workload. GPGPU was used to perform acceleration/parallelization of this computation.  On the hardware side, GPUs are widely available, cheap and scalable and on the software side GPU programming has become widely accessible, portable and supported by various compilers. 

We used NVIDIA's compute unified device architecture (CUDA) platform~\cite{cuda}. Optimization of a CUDA program depends somewhat on the GPU architecture used. We report results run on a consumer-level graphics card with the NVIDIA GeForce GTX 670 GPU and the higher-end NVIDIA Tesla K20 GPU. Both GPUs are based on the Kepler architecture. GeForce GTX 670 has peak theoretical double floating point performance of 0.19 TFlops whereas Tesla K20 has peak theoretical double floating point performance of 1.17 TFlops.  While GPU threads are plentiful and can be launched and terminated with minimal overhead, performance of parallelized flow map (trajectory) computation is primarily limited by memory bandwidth. 

For the GPU implementation only registers, global memory and constant memory were utilized. Registers were used by individual threads to store local variables in specific interpolation and integration kernels discussed below, and were reused in each kernel to the greatest extent possible. Since constant memory is read only and limited to 64kb but can be accessed by all threads, we used this for mesh parameters and other constants. Tracer and velocity field arrays were always stored on global memory.  

Each streaming multiprocessor (SMX) can run up to 16 resident blocks or 2048 resident threads concurrently for the architecture we used. A common target in optimizing the code was to make sure every streaming processor had roughly 2048 resident threads running at any given time. The number of registers available per SMX limits the number of threads that can reside on a SMX. Since the number of registers were limited to 64K per multiprocessor, this implied that every thread needed to use only 31 registers for 100\% occupancy. We developed our kernels, as discussed below, to keep register usage nominally within this bound.  

\subsection{Implementation}

\begin{figure}[h]
\centering
\includegraphics[width=\textwidth]{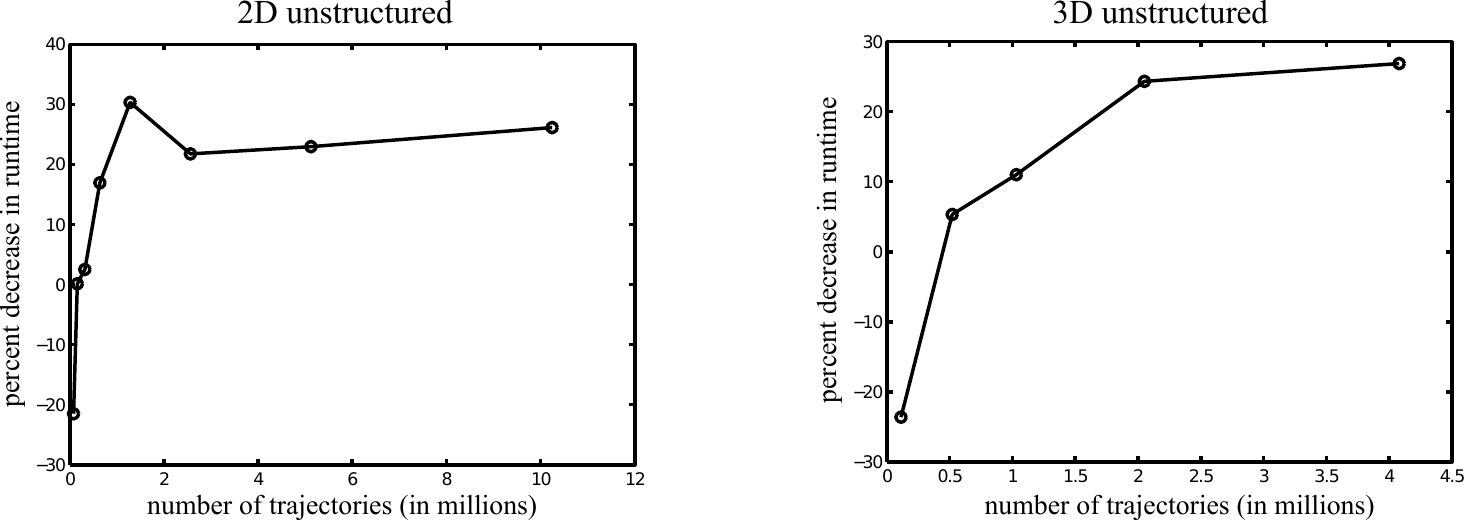}
\caption{Runtime comparison between a single-kernel RK4 time step implementation and a split-kernel RK4 time step implementation. Graphs plot the percent decrease in runtime of the split kernel implementation from that of the single kernel implementation. Runs were performed on both the 2D and 3D unstructured data examples described below.}
\label{fig:Kernel_comparison}
\end{figure}

For the purpose of explaining the implementation and performance results we focus discussion on the 4th order explicit Runge-Kutta integration method (RK4). To decrease register usage and hence improve GPU utilization, we broke up the integration kernel into smaller kernels. For example, RK4 requires the vector field to be evaluated at 4 different locations in space-time for a single update step. These interpolations are done as separate kernels. Furthermore, for unstructured grid data each interpolation requires a cell-search, \ie determination of which velocity grid cell the interpolation point is located.  Therefore, for each integration step, 8 kernels are run; 4 kernels to evaluate the velocity field, and each of these requires a preceding cell-search kernel call. This strategy involves the use of a large number of threads that are short-lived, which are what GPUs are designed to handle. Using this strategy also enables the CG tensor to be evaluated at any time step during the integration interval $(t_0, t_f)$ with minimal effort. This can be advantageous if the CG tensor is to be evaluated early for trajectories that leave the fluid domain before the full integration interval is reached. This is common with velocity data coming from modeling or measurement over a truncated domain. But moreover, some LCS detection strategies require certain criteria to be evaluated on the fly as integration proceeds. This need also motivated updating tracers synchronously for each integration step, which is consistent with maintaining efficient data flow in our pipeline, Sec.~\ref{sec:pipeline}. That is, the pipeline is most efficient if all upstream requests require the same velocity data loaded in memory. We note however that velocity data is loaded into GPU memory only every time a new velocity data file is needed, not every integration time step. Figure \ref{fig:Kernel_comparison} shows the runtime comparison of the single-kernel code and split-kernel code. It can be seen that split-kernel strategy executes in roughly 30\% less time for all by very small seed grid sizes. 

We use an efficient local cell search strategy for unstructured grids~\cite{ShaddenAstorinoGerbeau10}. We have demonstrated that this method outperforms other cell search methods we have tested such as the structured auxiliary mesh approach, and VTK's Kd-tree and Oct-tree methods. This method is designed for tetrahedral grids (or triangular grids in 2D). Therefore, other grid topologies of velocity data that might need to be processed are tetrahedralized using built in VTK functionality as a preprocessing step in the \texttt{Reader} filter. This enables a single, efficient cell search algorithm to be used, which alleviates the need to develop and optimize different interpolation and integration kernels for different velocity grid topologies. However the tradeoff is that piece-wise linear representation using a tetrahedral grid may degrade interpolation accuracy when the native CFD grid has higher order elements. 

\subsection{Performance}

\begin{figure}[h]
\centering
\includegraphics[width=\textwidth]{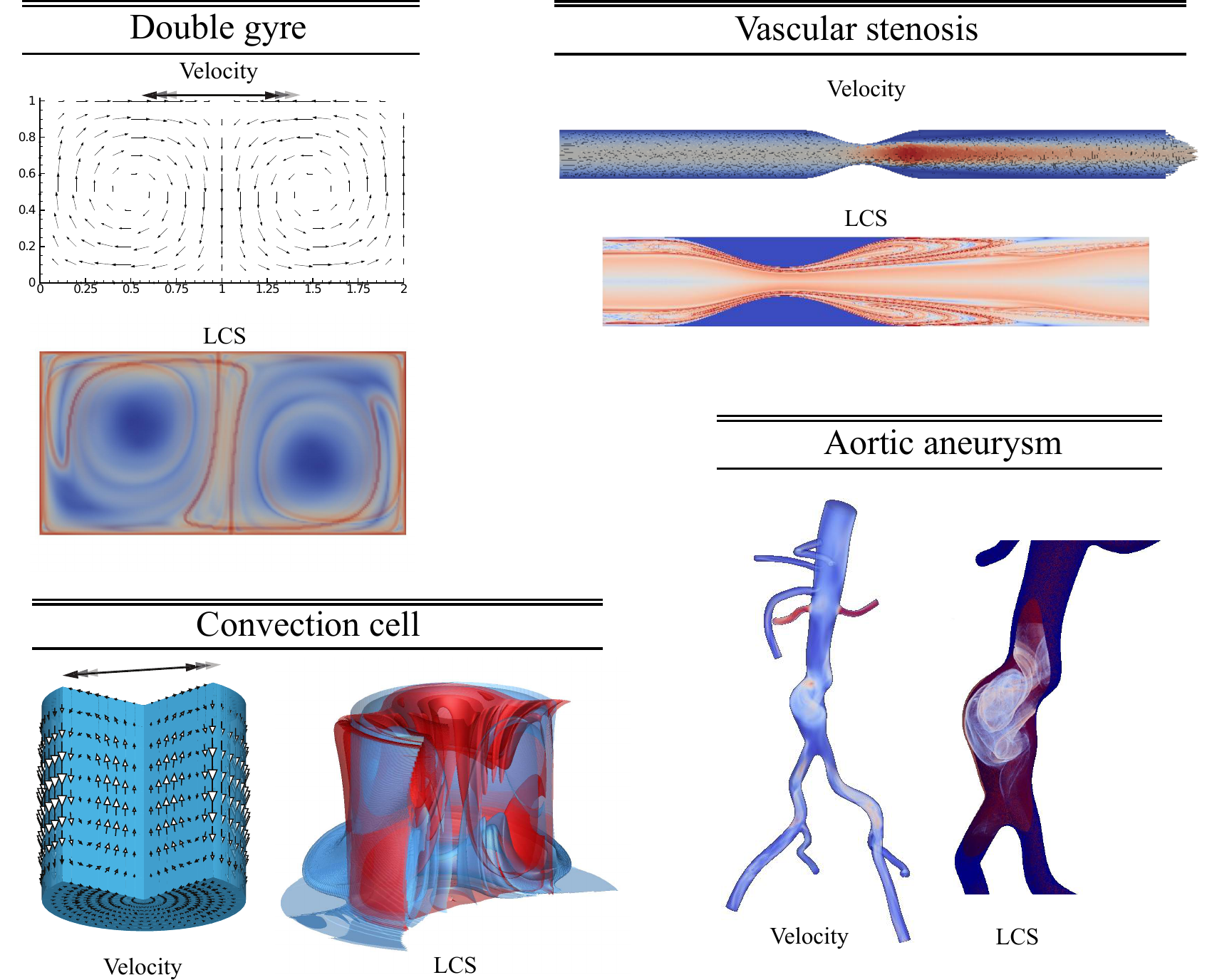}
\caption{Snapshots of the flow fields and LCS from the 4 applications used for performance testing, clockwise from upper-left: double gyre~\cite{ShaddenLekienMarsden05}; vascular stenosis~\cite{ShaddenHendabadi13}; abdominal aortic aneurysm~\cite{ArzaniShadden12}; Rayleigh-B\'enard convection~\cite{LekienShaddenMarsden07}.}
\label{fig:examples}
\end{figure}

Most of the processing time for the CG tensor field computation is spent inside the velocity field interpolation function. This function gets called one or more times each integration step for each trajectory update. Efficiency of this function is a main determinant of computational time. The interpolation function is mainly dependent on the grid topology of the velocity data, \eg interpolation on Cartesian grids differs in strategy and performance than interpolation on unstructured grids. Different kernels were developed to handle interpolation on different grid types. The efficient cell search algorithm for tetrahedralized unstructured grid data that is used readily facilitates spatial interpolation as described in~\cite{ShaddenAstorinoGerbeau10}. We present results for 4 different grid types considered: 2D Cartesian, 3D Cartesian, 2D unstructured, 3D unstructured. Specifically, the example applications include the double-gyre flow~\cite{ShaddenLekienMarsden05}, which has become a standard test case for LCS computations (2D Cartesian); the 3D Rayleigh-B\'enard convection cell~\cite{LekienShaddenMarsden07} (3D Cartesian); a coronary stenosis model~\cite{ShaddenHendabadi13} (2D unstructured); and an abdominal aortic aneurysm (AAA) model~\cite{ArzaniShadden12} (3D unstructured). These examples are shown in Fig.~\ref{fig:examples}.

\begin{figure}[h]
\centering
\includegraphics[width=\textwidth]{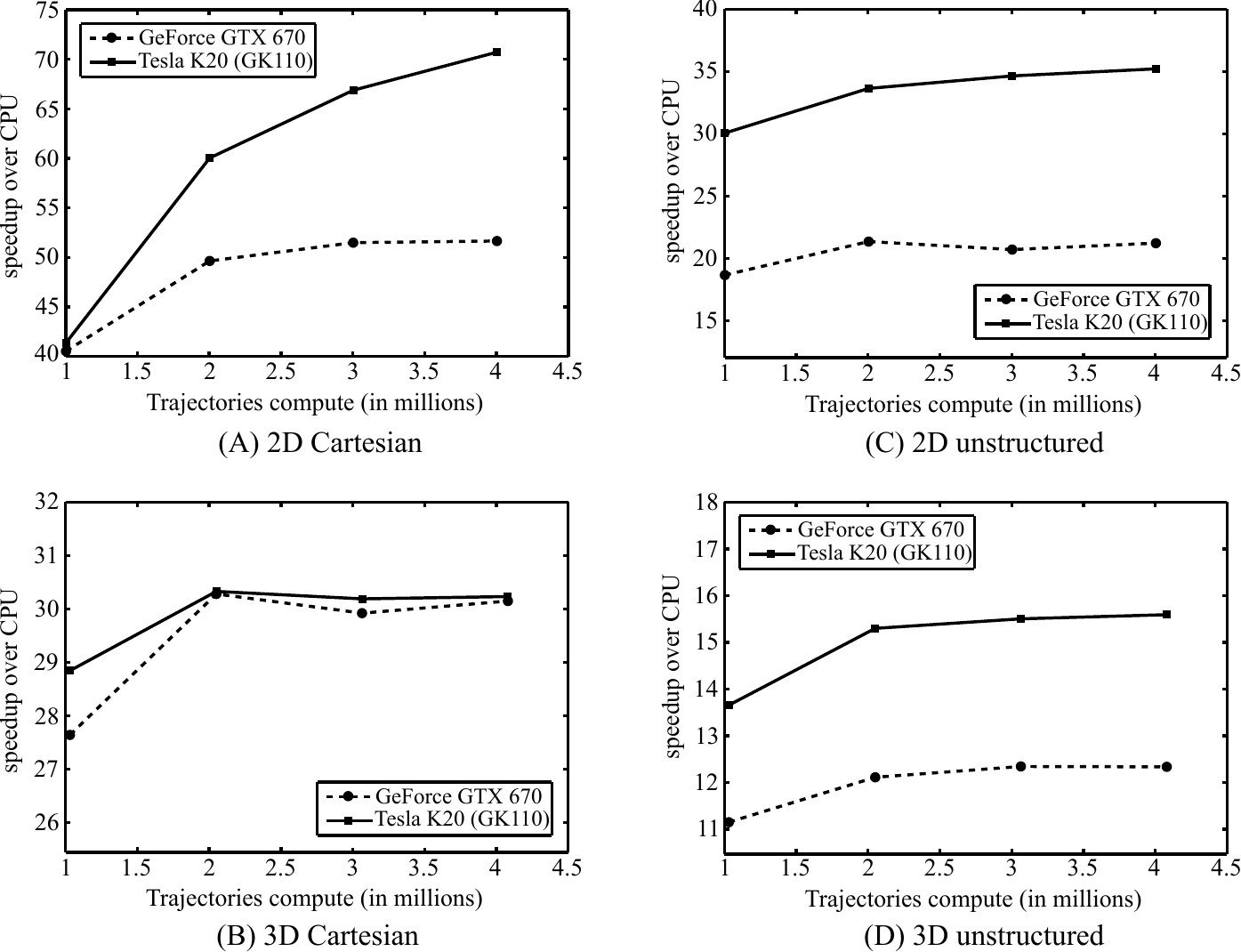}
\caption{Speedup of particle trajectory computation on a GeForce GTX 670 GPU and a Tesla K20 GPU compared to serial processing on an Intel Core i7-3770 CPU. All computations were performed using double precision for floating point variables.}
\label{fig:performance}
\end{figure}

Fig.~\ref{fig:performance} plots the performance results from the GeForce and Tesla GPUs. The values plotted are the runtimes on the respective GPU normalized by the serial CPU runtime, which gives the speedup over serial CPU processing. The CPU used was an Intel Core i7-3770 Ivy Bridge 3.5GHz processor and the implementation was FlowVC~\cite{FlowVC}, which was written in C. We observed up to roughly 70x speedup for computations on the 2D Cartesian data down to roughly 12x speed up for 3D unstructured grid data. As expected, the Tesla K20 yields higher speedup due to a higher number of streaming multiprocessors. The AAA velocity data was specified on an unstructured tetrahedral grid of 1.01 million elements. Unstructured velocity grids up to several million elements were tested and yielded similar speedup results as the AAA model.

We note that we achieved significant performance improvement by properly coalescing tracer grid arrays, which is common practice, whereby nearby threads processed tracer data coalesced in memory. We did not notice any improvement of performance by sorting velocity field data arrays. This is because as the trajectories of the seed grid evolve, trajectories with nearby indices in memory (and hence locally processed on the GPU) do not necessarily require velocity data that is localized in space.  A recent solution to this problem was proposed by~\cite{ChenHart}, which reorganizes particles into spatially coherent bundles as they are advected to improve memory coherence and shared-memory GPU performance.

\subsection{Verification}

Computations performed on the GPU were verified against existing CPU code that has been extensively used and tested, FlowVC. We present verification data from the 2D Cartesian and 3D unstructured grid applications, as these cases represent the extrema in results from the 4 examples considered. For each application we released several thousand tracers in the domain and integrated their trajectories for the nominal minimum time needed for LCS computation. Specifically, we chose this time based on the the AAA model, and scaled the other integration times by the nominal edge size $x_l$ divided by the mean velocity magnitude $\langle \bu(\bx,t) \rangle$ at peak flow. This defines a characteristic time scale similar to use of the CFL number. This ensured that all cases were integrated over roughly the same number of elements in their respective domains. We performed these computations using the GPU and CPU codes on the GeForce GTX 670 and Intel Core i7-3770. The average error was defined by computing of the $L_2$ norm of the differences in tracer locations over time between the two runs, and averaging over all tracers as follows
$$
e_D(t) = \frac{1}{N} \sum_{i=1}^N \left\| \mathbf{x}_{i}^{DG}(t) - \mathbf{x}_{i}^{DC}(t) \right\| \;,
$$ 
where $N$ is the number of trajectories,  $\mathbf{x}_{i}^{DG}(t)$ is the trajectory of tracer $i$ computed on the GPU using double precision floating point numbers, $\mathbf{x}_{i}^{DC}(t)$ is the trajectory of tracer $i$ computed on the CPU using double precision floating point numbers, and $\mathbf{x}_{i}^{DG}(0) = \mathbf{x}_{i}^{DC}(0)$. In all applications the error stayed below $1\times10^{-10}$, indicating that both GPU and CPU implementations were performing equivalent tracer trajectory computation. 

\begin{figure}[h]
\centering
\includegraphics[width=\textwidth]{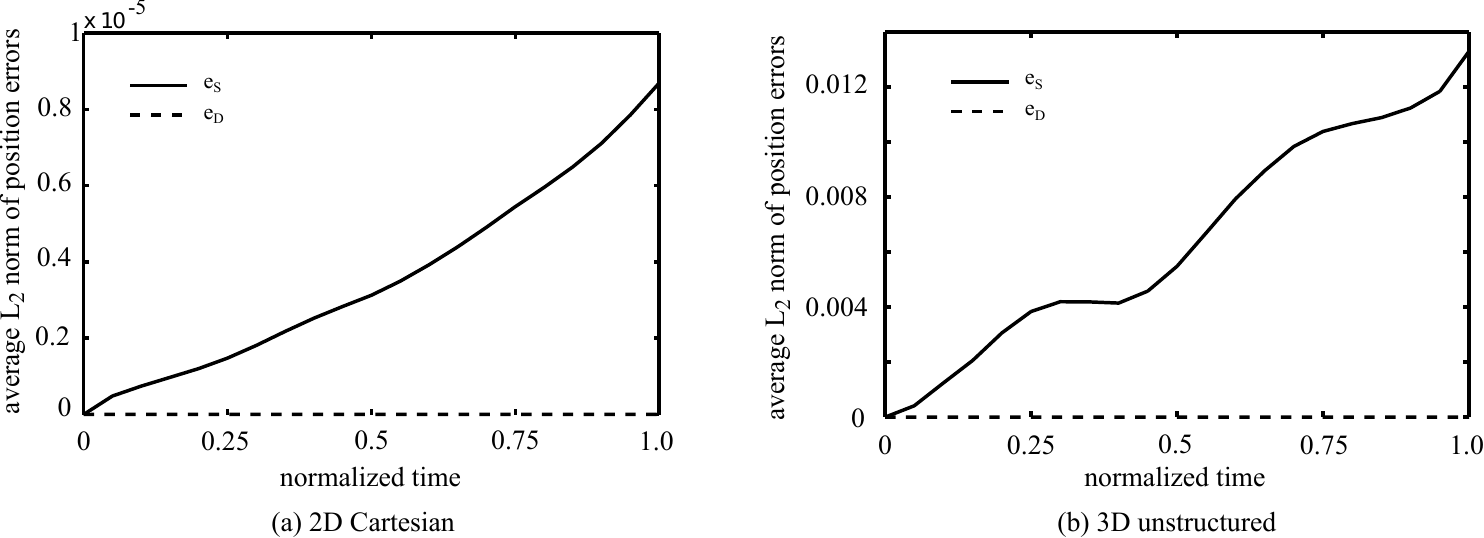}
\caption{Error between single precision (solid line) and double precision (dashed line) GPU computations compared to a double precision CPU computations versus integration time for the double gyre flow (left) and AAA flow (right). Integration times have been normalized and the value of 1.0 approximately represents the nominal minimum integration time needed to compute LCS for each application.}
\label{fig:errors}
\end{figure}

Single precision calculation can be several times faster than the double precision calculation, however accuracy may be unacceptable. The results generated using a single precision floating point GPU implementation were compared against the double precision CPU results. As described above, the average $L_2$ norm of the error in trajectories of several thousand tracers was computed from a single precision GPU run and a double precision CPU run as
$$
e_S(t) = \frac{1}{N} \sum_{i=1}^N \left\| \mathbf{x}_{i}^{SG}(t) - \mathbf{x}_{i}^{DC}(t) \right\| \;,
$$
where $\bx_i^{SG}(t)$ is the trajectory of tracer $i$ computed on the GPU using single precision for floating point variables. For the 2D Cartesian data, we noticed fairly acceptable errors in tracer trajectories and subsequent LCS computation. However, for the more complex 3D unstructured grid data, we notice unacceptable degradation in accuracy, \eg to a point where noticeable degradation of the FTLE field occurred. The errors $e_d(t)$ and $e_s(t)$ are plotted against integration time for these two cases in Fig.~\ref{fig:errors} using the NVIDIA GeForce GTX 670 GPU.

Because the plots in Fig.~\ref{fig:errors} represent averages over many tracers, some tracer trajectories may deviate to far greater extent than the mean values shown. Indeed, in the AAA flow, some tracers were advected to different arterial branches based on these errors, which has important consequences for that application. Not surprisingly, errors can be worse near LCS due to inherent sensitivity to initial conditions at these locations, which can be problematic for accurate LCS detection. In addition, while we consider the double precision CPU results as a baseline for comparison, this does not imply that these results represent the ``true'' trajectories. The double precision CPU computations are subject to normal truncation and round-off errors. However, since double precision computation represents the de facto method for minimizing numerical error, and since no application considered nor of practical importance has a closed-form analytic solution, this was deemed an appropriate baseline for comparison.      

\section{Discussion} \label{sec:discussion}

We have developed a modular pipeline for LCS computation that is capable of loading a wide variety of fluid mechanics data, and that can be easily interfaced with ParaView for visualization of LCS computational results alongside other flow visualization tools. This pipeline was designed to be modular and flexible so that modifications and additions can be made with minimal effort. 

As the LCS method has gained considerable popularity in the past few year, there have been several developments to improve computation and visualization of LCS. Strategies have been devised so that tracer seeding can be adapted to cluster sampling of the flow map near LCS for improved detection~\cite{GarthETAL07, SadloPeikert07, SadloRigazziPeikert10, LekienRoss10}. Also, since vector field interpolation is computationally intensive, strategies have also been developed to locally approximate the flow map~\cite{BruntonRowley10}, which can be beneficial when time series of tracer grids are considered. As well, trajectory computations obtained in a hierarchical manner have been considered for efficient FTLE computation~\cite{HlawatschSadloWeiskopf11}. We note that sampling the flow map (gradient) needs to be highly resolved in space but not necessarily time for LCS computations. That is, sampling in space is driven by LCS identification methods; mainly computation of the flow map gradient or subsequently the CG tensor field.  Sampling time is done primarily for visualizing the evolution of the structures. Regardless of the strategy of sampling or interpolating the flow map or its derivative over the fluid domain, this process can be parallelized since advection of tracers can be performed for the most part independently.  Because of this high degree of parallelism, and the fact the LCS are typically used for desktop postprocessing and flow visualization, many-core implementation on a single workstation is desirable, as opposed to a visualization cluster, which are less accessible. 

Jimenez and Vankerschaver~\cite{JimenezVankerschaver09} discussed the computation and visualization of FTLE by GPGPU using CUDA. They also made their source code publicly available~\cite{JimenezVankerschaver-software}. Their implementation was, by their own admission, naive since it only could handle FTLE computation for analytically defined vector fields. This greatly reduced memory accesses that, as discussed above, are the bottleneck in computing FTLE fields using GPU processing in practical applications. Garth, et al.,~\cite{GarthETAL07} and Hlawatsch, et al.~\cite{HlawatschSadloWeiskopf11} leveraged GPU processing for FTLE computation as well, though these papers did not discuss details of their implementation or performance results as this was not a main focus. The recent paper by Conti, et al.~\cite{ContiRossinelliRossinelli12} described FTLE computation for bluff body flows using  OpenCL, which allows implementation on mixed architectures, \eg AMD's accelerated processing units.  Their implementation was specialized to remeshed vortex methods for bluff body flows that involve mesh-particle interpolations previously tailored for GPU implementation~\cite{RossinelliContiKoumoutsakos11}. They were able to achieve around one order magnitude speedup compared to serial CPU implementation, similar to what we observed here. However, their application was specialized to a particular flow problem. Most LCS computation is performed as a post processing step on fluid velocity field data. Since this is the most common and general scenario, it was the one which we developed our framework around, consistent with the design specification of our LCS pipeline. 

We considered application to 3D flow and flow on unstructured grids, as~\cite{JimenezVankerschaver09, ContiRossinelliRossinelli12} reported performance results for 2D flows on structured grids. As shown and discussed above, we have run this implementation to integrate millions of tracers on velocity grids with several million element. This represents a reasonable limit for most LCS applications. For significantly larger velocity field grids, or tracer grids, one will overflow the global memory for the GPU.  We generally noticed peak performance when kernels were kept within register memory bounds. This required each trajectory update to be divided into a series of kernels. Therefore, kernels were very short. Because of the way data is processed through our pipeline, we perform synchronous integration between the tracers. We believe this has advantages for LCS detection as well. For example, in truncated domains, which represent the vast majority of fluid mechanics data, tracers leave the domain before the finite time interval being considered for flow map computation. But perhaps more importantly, one may want to have access to the CG tensor at various times. Most LCS criteria are a one parameter family dependent on the chosen integration window. Using one integration window may identify a manifold as an LCS, but another integration window may not. Similarly, a manifold may satisfy one LCS criterion but not the other for a particular integration time. Therefore, having access to this information enables implementation of methods that may depend on how strain rate or direction changes over the integration time parameter. Or alternatively, schemes that adaptively sample the flow map based on CG tensor information may need sequential access to this information. 

While this pipeline can function as a stand alone program, it can also be compiled as libraries to be used with ParaView. Such integration provides a natural platform for visualizing not only LCS computations, but also integrating these results with other existing flow visualization tools provided by this programs, and also GPU-based visualization techniques available. Indeed, LCS as a collection of codimension one surfaces are rarely useful in understanding the flow. Knowledge of these manifolds must be integrated with other knowledge of the flow to gain fundamental insight into the flow topology, and these platforms provide significant capability in this regard.

\begin{acknowledgement}
This work was supported by the National Science Foundation, award number 1047963. 
\end{acknowledgement}

\bibliographystyle{spmpsci}

\end{document}